\documentclass[aps,prd,onecolumn,groupedaddress,showpacs,nofootinbib,amssymb]{revtex4}
\usepackage[T1]{fontenc}
\usepackage[latin1]{inputenc}
\usepackage{graphicx}
\usepackage[english]{babel}
\usepackage{amsmath}
\usepackage{amssymb}
\usepackage{amsfonts}
\usepackage{bm}

\begin{document}

\def\pp{{\, \mid \hskip -1.5mm =}}
\def\cL{{\cal L}}
\def\be{\begin{equation}}
\def\ee{\end{equation}}
\def\bea{\begin{eqnarray}}
\def\eea{\end{eqnarray}}
\def\beq{\begin{eqnarray}}
\def\eeq{\end{eqnarray}}
\def\tr{{\rm tr}\, }
\def\nn{\nonumber \\}
\def\e{{\rm e}}
\title{INVERSE PROBLEM - RECONSTRUCTION OF DARK ENERGY MODELS}

\author{\footnotesize SHIN'ICHI NOJIRI}
\affiliation{Department of Physics, Nagoya University, Nagoya 464-8602, Japan}

\begin{abstract}

We review how we can construct the gravity models which 
reproduces the arbitrary development of the universe. 
We consider the reconstruction in the Einstein gravity coupled with generalized perfect fluid, 
scalar-Einstein gravity, scalar-Einstein-Gauss-Bonnet gravity, 
Einstein-$F(\mathcal{G})$-gravity, and $F(R)$-gravity. 
Very explicit formulas are given to reconstruct the models, which could be used when we find 
the detailed data of the development of the universe by future observations. 
Especially we find the formulas using e-foldings, which has a direct relation with observed redshift. 
As long as we observe the time development of the Hubble rate $H$, there exists a variety of models 
describing the arbitrary development of universe. 

\end{abstract}

\pacs{95.36.+x, 98.80.Cq, 04.50.Kd, 11.10.Kk, 11.25.-w}

\maketitle

\section{Introduction}

We now believe that there was an epoch called inflation in the early universe. 
In the epoch, the expansion of the universe was rapidly accelerating. 
On the other hand, several observations tell that the expansion of the present universe is 
also accelerating \cite{WMAP1,Komatsu:2008hk,SN1}. 
There were many epochs between the inflation and the late time acceleration, like, reheating, 
transparent to radiation, matter dominance, etc. 
In order to explain the complicated development of the universe, especially the inflation and/or 
the accelerating expansion of the present universe, several models of extended or modified gravity 
were proposed. 
Usually we start from a theory, which is usually given by action or something, and solve the dynamics. 
In this report, however, we consider the inverse problem, that is, 
we investigate the reconstruction of gravity theories, that is, 
we show how we can construct the gravity models which 
reproduces the complicated development of the universe 
when we find the development by the observations etc. 
We show the reconstruction in the models of the Einstein gravity 
with generalized perfect fluid, scalar-Einstein gravity, scalar-Einstein-Gauss-Bonnet gravity, 
Einstein-$F(\mathcal{G})$-gravity, $F(R)$-gravity. 
Just for simplicity, we neglect the contributions from matters although it is easy to
include them. 
For general review of reconstruction in modified gravity, see \cite{Sahni:2006pa,Nojiri:2006be}.
This kind of reconstruction may be used also for spherical symmetric solution like 
black holes (see, for example, \cite{Nojiri:2009uu}). 

In this report, we assume the Friedman-Robertson-Walker (FRW) universe with flat spacial part:
\be
\label{ma1}
ds^2= -dt^2 + a(t)^2\sum_{i=1,2,3} \left(dx^i\right)^2\, .
\ee
Here $a(t)$ is the scale factor at cosmological time $t$. 

We often use the equation of state (EoS) parameter for a perfect fluid defined by
\be
\label{ma4}
w = \frac{p}{\rho}\, .
\ee
Here $\rho$ is the energy density and $p$ is the pressure of the perfect fluid. 
In the Einstein gravity, by using the FRW equations, 
\be
\label{ma2}
\rho_\mathrm{total} = \frac{3}{\kappa^2}H^2\, ,\quad p_\mathrm{total}
=-\frac{1}{\kappa^2}\left(2\dot H + 3H^2\right)\, ,
\ee
we find
\be
\label{ma5}
w_\mathrm{total} = \frac{p_\mathrm{total}}{\rho_\mathrm{total}} = - 1 - \frac{2\dot H}{3H^2}\, .
\ee
Here the Hubble rate $H$ is defined by $H\equiv \dot a/a$. 
In (\ref{ma5}), $\rho_\mathrm{total}$ and $p_\mathrm{total}$ are total energy density 
and total pressure, respectively. 
In the following, even for modified or extended gravities, we define effective EoS parameter by
\be
\label{ma6}
w_\mathrm{eff} = - 1 - \frac{2\dot H}{3H^2}\, .
\ee

\section{Generalized Equation of State}

First we consider the Einstein Gravity coupled with generalized perfect fluid \cite{Nojiri:2005sr}. 
The FRW equations (\ref{ma2}) tell that 
if we consider the perfect fluid with the generalized equation of state
\be
\label{ma3}
p=-\rho -
\frac{2}{\kappa^2}f'\left(f^{-1}\left(\kappa\sqrt{\frac{\rho}{3}}\right)\right) \, ,
\ee
we find the solution of the FRW equations by $H=f(t)$. 
Here $f(t)$ can be an arbitrary function of the cosmological time $t$. 
Therefore we find the arbitrary development of universe given by $H=f(t)$ can be realized 
by the perfect fluid whose equation of state is given by (\ref{ma3}). 

It is often convenient to use redshift $z$ instead of cosmological time $t$ since the redshift has 
direct relation with observations. The redshift is defined by
\be
\label{ma14a}
a(t) = \frac{a\left(t_0\right)}{1+z} = \e^{N - N_0}\, .
\ee
Here $t_0$ is the cosmological time of the present universe, $N_0$ could be an arbitrary constant, 
and $N$ is called as e-folding and directly related with the redshift $z$. 
We now consider the reconstruction by using $N$ instead of the cosmological time $t$. 
Then the second equation in (\ref{ma2}) can be rewritten as
\be
\label{ma2b}
p_\mathrm{total} = - \frac{1}{\kappa^2} \left(2H H' + 3H^2\right)\, .
\ee
Here $'$ expresses the derivative with respect to $N$: $'\equiv d/dN$. 
Then if we consider the perfect fluid with the generalized equation of state
\be
\label{ma3b}
p=-\rho -
\frac{2}{\kappa}\tilde f'\left(\tilde f^{-1}\left(\kappa\sqrt{\frac{\rho}{3}}\right)\right) 
\sqrt{\frac{\rho}{3}}\, , 
\ee
we find the solution of the FRW equations by $H=\tilde f(N)$. 

\section{Scalar-Einstein gravity}

In this section, we consider the reconstruction of scalar-Einstein gravity (or scalar-tensor theory), 
whose action can be written as
\be
\label{ma7}
S=\int d^4 x \sqrt{-g}\left\{\frac{1}{2\kappa^2}R - \frac{1}{2}\omega(\phi)\partial_\mu \phi 
\partial^\mu\phi - V(\phi)\right\}\, .
\ee
Here $\omega(\phi)$, $V(\phi)$ are functions of the scalar field $\phi$. 
The function $\omega(\phi)$ is not relevant and can be absorbed into the redefinition of 
the scalar field $\phi$. In fact if we redefine the scalar field $\phi$ by 
\be
\label{ma13}
\varphi \equiv \int^\phi d\phi \sqrt{\left|\omega(\phi)\right|} \, ,
\ee
the kinetic term of the scalar field in the action (\ref{ma7}) has the following form:
\be
\label{ma13b}
 - \omega(\phi) \partial_\mu \phi \partial^\mu\phi 
 = \left\{ \begin{array}{ll} 
 - \partial_\mu \varphi \partial^\mu\varphi & \mbox{when $\omega(\phi) > 0$} \\
\partial_\mu \varphi \partial^\mu\varphi & \mbox{when $\omega(\phi) < 0$} 
\end{array} \right. \, .
\ee
The case of $\omega(\phi) > 0$ corresponds to the quintessence or non-phantom scalar field but 
the case of $\omega(\phi) < 0$ corresponds to the phantom scalar. 
Although $\omega(\phi)$ can be absorbed into the redefinition of the scalar field, 
we keep $\omega(\phi)$ for the later convenience. 

\subsection{Formulation by using the cosmological time $t$} 

This section is based on the works \cite{Nojiri:2005pu,Capozziello:2005tf}. 

For the action (\ref{ma7}), in the FRW equations (\ref{ma2}), the energy density and the pressure 
have the following forms:
\be
\label{ma8}
\rho = \frac{1}{2}\omega(\phi){\dot \phi}^2 + V(\phi)\, ,\quad 
p = \frac{1}{2}\omega(\phi){\dot \phi}^2 - V(\phi)\, .
\ee
Then we find
\be
\label{ma9}
\omega(\phi) {\dot \phi}^2 = - \frac{2}{\kappa^2}\dot H\, ,\quad 
V(\phi)=\frac{1}{\kappa^2}\left(3H^2 + \dot H\right)\, .
\ee
If we assume $\omega(\phi)$ and $V(\phi)$ are given by a single function $f(\phi)$, as follows, 
\be
\label{ma10}
\omega(\phi)=- \frac{2}{\kappa^2}f'(\phi)\, ,\quad 
V(\phi)=\frac{1}{\kappa^2}\left(3f(\phi)^2 + f'(\phi)\right)\, ,
\ee
we find that the exact solution of the FRW equations (\ref{ma2}) with (\ref{ma8}) (when we neglect the 
contribution from the matter) has the following form: 
\be
\label{ma11}
\phi=t\, ,\quad H=f(t)\, .
\ee
We can confirm that the equation given by the variation of $\phi$  
\be
\label{ma12}
0=\omega(\phi)\ddot \phi + \frac{1}{2}\omega'(\phi){\dot\phi}^2 + 3H\omega(\phi)\dot\phi 
+ V'(\phi)\, ,
\ee
is also satisfied by the solution (\ref{ma11}). 
Then the arbitrary development of universe expressed by $H=f(t)$ can be realized by a appropriate choice 
of $\omega(\phi)$ and  $V(\phi)$. 
In other words, if we find the development of the universe, we can find the corresponding 
scalar-Einstein gravity. 

As mentioned in (\ref{ma13}) and (\ref{ma13b}), $\omega(\phi)$ can be absorbed into the redefinition of 
the scalar field $\phi$.   
By keeping $\omega(\phi)$, however, we can construct a model which shows the smooth 
transition from non-phantom phase to phantom phase. 

\subsection{Formulation by using e-folding $N$}

In this subsection, we consider the formulation by using the e-folding $N$ in (\ref{ma14a}). 
The FRW equations  (\ref{ma2}) with (\ref{ma8}) (when we neglect the 
contribution from the matter) are given in terms of $N$, as follows, 
\be
\label{ma15}
\frac{3}{\kappa^2} H^2 = \frac{H^2 \omega\left(\phi\right){\phi'}^2 }{2} + V\left(\phi\right)\, ,\quad 
 - \frac{1}{\kappa^2} \left(2H H' + 3 H^2 \right)  
= \frac{H^2 \omega\left(\phi\right){\phi'}^2 }{2} - V\left(\phi\right)\, .
\ee
We now identify $\phi=N$, then we find 
\be
\label{ma16}
\omega(\phi) = - \frac{2H'}{\kappa^2 H}\, ,\quad
V\left(\phi\right) = \frac{1}{\kappa^2}\left( H H' + 3 H^2 \right)\, .
\ee
The above equations in (\ref{ma16}) tells that if we consider the model
\be
\label{ma17}
\omega(\phi) = - \frac{2f'\left(\phi\right)}{\kappa^2 \tilde f\left(\phi\right)}\, ,\quad 
V\left(\phi\right) = \frac{1}{\kappa^2}\left( \tilde f\left(\phi\right) \tilde f'\left(\phi\right) 
+ 3 \tilde f\left(\phi\right)^2 \right)\, ,
\ee
which is given in term of a single function $\tilde f$, then the exact solution is given by 
$H\left(N\right) = f\left(N \right)$. 

We now consider an example by using the following function $\tilde f$, 
\be
\label{ma18}
\tilde f(\phi) = H_1 \phi^{-\gamma}\, .
\ee
Here $H_1$ and $\gamma$ are positive constants. Then we find
\be
\label{ma18b}
\omega(\phi) = \frac{2\gamma}{\kappa^2 \phi}\, ,\quad
V\left(\phi\right) = \frac{H_1^2}{\kappa^2}\left( - \gamma \phi^{-2\gamma -1}  
+ 3 \phi^{-2\gamma} \right)\, ,
\ee
and the exact solution is given by $H = H_1 N^{-\gamma}$.
If we redefine $\phi$ by $\phi\to \varphi = \frac{2 \sqrt{2\gamma\phi}}{\kappa}$, 
the kinetic term becomes canonical and the potential is given by
\be
\label{ma20}
V\left(\varphi\right) = \frac{H_1^2}{\kappa^2}\left\{ - \gamma 
\left(\frac{\kappa^2}{8\gamma}\right)^{- 2\gamma - 1}  \varphi^{-4\gamma - 2}  
+ 3 \left(\frac{\kappa^2}{8\gamma}\right)^{- 2\gamma}  \varphi^{-4\gamma} \right\}\, .
\ee
In the late time (large $N$), only the last term is relevant. 
Conversely, in the late time, if the potential is given by 
$V \propto \varphi^{-4\gamma}$, the solution has the form of $H \propto N^{-\gamma}$. 
Then the effective EoS parameter (\ref{ma6}) is given by
\be
\label{ma19}
w_\mathrm{eff} = - 1 + \frac{2\gamma}{3N} \, ,
\ee
which goes to $-1$ when $N$ goes to infinity. 

Since we naively find $C \equiv \frac{\mbox{Plank scale}}{H_0} \sim 10^{61} \sim \e^{140}$, 
if $\frac{H_1}{H_0} \sim C$, we find $N \sim \ln C$ and therefore $\gamma \sim 28$. 
Then the fine tuning problem might be relaxed but in this case we obtain $w_\mathrm{eff}\sim - 0.87$, 
which is little bit greater than the value obtained 
from the observation $-0.14 < 1+w < 0.12$ \cite{Komatsu:2008hk}.

When we consider the case that $\tilde C \equiv \frac{\mbox{Weak scale}}{H_0} \sim 10^{44} \sim \e^{101}$, 
if $\frac{H_1}{H_0} \sim \tilde C$ we find $N \sim \ln \tilde C$ and $\gamma \sim 21$. 

\section{Scalar-Einstein-Gauss-Bonnet gravity}

In this section, we consider the reconstruction of the scalar-Einstein-Gauss-Bonnet gravity\footnote{For pioneering work 
on the scalar-Einstein-Gauss-Bonnet gravity, see \cite{Boulware:1986dr}. }  
based on \cite{Nojiri:2005vv,Nojiri:2006je}. 

The starting action has the following form: 
\be
\label{ma22}
S=\int d^4 x \sqrt{-g}\left[ \frac{R}{2\kappa^2} - \frac{1}{2}\partial_\mu \phi \partial^\mu \phi 
 - V(\phi) - \xi(\phi) \mathcal{G}\right]\, .
\ee
Here $\mathcal{G}$ is the Gauss-Bonnet invariant, defined by 
$\mathcal{G}\equiv R^2 - 4 R_{\mu\nu} R^{\mu\nu} + R_{\mu\nu\rho\sigma} R^{\mu\nu\rho\sigma}$. 
Different from the case of the scalar-Einstein gravity action in (\ref{ma7}), we restrict the kinetic term of 
the scalar field to be canonical. 
The scalar-Gauss-Bonnet gravity appears as a stringy correction to the Einstein gravity 
but in this section, we do not restrict the model to the stringy one. 

Under the variation of the action (\ref{ma22}) over the metric $g_{\mu\nu}$, we obtain 
\bea
\label{ma23}
&& 0= \frac{1}{\kappa^2}\left(- R^{\mu\nu} + \frac{1}{2} g^{\mu\nu} R\right)
+ \frac{1}{2}\partial^\mu \phi \partial^\nu \phi
 - \frac{1}{4}g^{\mu\nu} \partial_\rho \phi \partial^\rho \phi 
+ \frac{1}{2}g^{\mu\nu}\left( - V(\phi) + \xi(\phi) \mathcal{G} \right) \nn
&& -2 \xi(\phi) R R^{\mu\nu} - 4\xi(\phi)R^\mu_{\ \rho} R^{\nu\rho} 
 -2 \xi(\phi) R^{\mu\rho\sigma\tau}R^\nu_{\ \rho\sigma\tau}
+4 \xi(\phi) R^{\mu\rho\nu\sigma}R_{\rho\sigma} \nn
&& + 2 \left( \nabla^\mu \nabla^\nu \xi(\phi)\right)R 
 - 2 g^{\mu\nu} \left( \nabla^2\xi(\phi)\right)R 
 - 4 \left( \nabla_\rho \nabla^\mu \xi(\phi)\right)R^{\nu\rho} 
 - 4 \left( \nabla_\rho \nabla^\nu \xi(\phi)\right)R^{\mu\rho} \nn
&& + 4 \left( \nabla^2 \xi(\phi) \right)R^{\mu\nu}
+ 4g^{\mu\nu} \left( \nabla_{\rho} \nabla_\sigma \xi(\phi) \right) R^{\rho\sigma}
+ 4 \left(\nabla_\rho \nabla_\sigma \xi(\phi) \right) R^{\mu\rho\nu\sigma} \, .
\eea
In the expression of (\ref{ma23}), there do not appear the derivatives of curvature like $\nabla R$ 
and therefore the derivatives higher than two do not appear, which could be contrasted with a general 
$\alpha R^2 + \beta R_{\mu\nu}R^{\mu\nu} + \gamma R_{\mu\nu\rho\sigma}R^{\mu\nu\rho\sigma}$-gravity, 
where fourth derivatives of $g_{\mu\nu}$ appear. 
Then if we consider the classical theory, if we specify the values of $g_{\mu\nu}$ and $\dot g_{\mu\nu}$ 
on a spacial surface as an initial condition, the time-development is uniquely determined. 
This situation is similar to the case in the classical mechanics, where we only need to specify the values of 
position and velocity of particle as initial conditions. 
In a general $\alpha R^2 + \beta R_{\mu\nu}R^{\mu\nu} + \gamma R_{\mu\nu\rho\sigma}R^{\mu\nu\rho\sigma}$-gravity, 
we need to specify the values of $\ddot g_{\mu\nu}$, $\dddot g_{\mu\nu}$ besides $g_{\mu\nu}$, $\dot g_{\mu\nu}$ 
so that we could obtain a unique time development. 
In the Einstein gravity, we only need to specify the value of $g_{\mu\nu}$, $\dot g_{\mu\nu}$ as initial conditions. 
Then the scalar-Gauss-Bonnet gravity is a natural extension of the Einstein gravity. 

In the FRW universe (\ref{ma1}), Eq.(\ref{ma23}) has the following forms:
\bea
\label{ma24}
0&=& - \frac{3}{\kappa^2}H^2 + \frac{1}{2}{\dot\phi}^2 + V(\phi) 
+ 24 H^3 \frac{d \xi(\phi(t))}{dt}\, ,\\
\label{GBany5}
0&=& \frac{1}{\kappa^2}\left(2\dot H + 3 H^2 \right) + \frac{1}{2}{\dot\phi}^2 - V(\phi) 
 - 8H^2 \frac{d^2 \xi(\phi(t))}{dt^2} - 16H \dot H \frac{d\xi(\phi(t))}{dt} \nn
&& - 16 H^3 \frac{d \xi(\phi(t))}{dt} \, .
\eea
On the other hand, by the variation of the scalar field, we obtain
\be
\label{ma24b}
0=\ddot \phi + 3H\dot \phi + V'(\phi) + \xi'(\phi) \mathcal{G}\, .
\ee
By combining (\ref{ma24}) and (\ref{GBany5}) and deleting $V(\phi)$, we obtain 
\bea
\label{ma25}
0 &=& \frac{2}{\kappa^2}\dot H + {\dot\phi}^2 - 8H^2 \frac{d^2 \xi(\phi(t))}{dt^2} 
 - 16 H\dot H \frac{d\xi(\phi(t))}{dt} + 8H^3 \frac{d\xi(\phi(t))}{dt} \nn
&=& \frac{2}{\kappa^2}\dot H + {\dot\phi}^2 
 - 8a\frac{d}{dt}\left(\frac{H^2}{a}\frac{d\xi(\phi(t))}{dt}\right)\, .
\eea
Then we obtain the following expressions of $\xi\left(\phi(t)\right)$ 
and $V\left(\phi(t)\right)$:
\bea
\label{ma26}
\xi(\phi(t)) &=& \frac{1}{8}\int^t dt_1 \frac{a(t_1)}{H(t_1)^2} \int^{t_1} \frac{dt_2}{a(t_2)}
\left(\frac{2}{\kappa^2}\dot H (t_2) + {\dot\phi(t_2)}^2 \right)\, , \nn
V(\phi(t)) &=& \frac{3}{\kappa^2}H(t)^2 - \frac{1}{2}{\dot\phi (t)}^2 
 - 3a(t) H(t) \int^t \frac{dt_1}{a(t_1)}
\left(\frac{2}{\kappa^2}\dot H (t_1) + {\dot\phi(t_1)}^2 \right)\, .
\eea
Therefore if we consider the model that $V(\phi)$ and $\xi(\phi)$ are given 
by adequate functions $g(t)$ and $f(\phi)$ as follows, 
\bea
\label{ma27}
V(\phi) &=& \frac{3}{\kappa^2}g'\left(f(\phi)\right)^2 - \frac{1}{2f'(\phi)^2} \nn
&& - 3g'\left(f(\phi)\right) \e^{g\left(f(\phi)\right)} \int^\phi d\phi_1 f'( \phi_1 )
\e^{-g\left(f(\phi_1)\right)} \left(\frac{2}{\kappa^2}g''\left(f(\phi_1)\right) 
+ \frac{1}{f'(\phi_1 )^2} \right)\, ,\nn
\xi(\phi) &=& \frac{1}{8}\int^\phi d\phi_1 \frac{f'(\phi_1) \e^{g\left(f(\phi_1)\right)} }{g'(\phi_1)^2} \nn
&& \times \int^{\phi_1} d\phi_2  f'(\phi_2) \e^{-g\left(f(\phi_2)\right)} 
\left(\frac{2}{\kappa^2}g''\left(f(\phi_2)\right) + \frac{1}{f'(\phi_2)^2} \right)\, ,
\eea
we find a solution of (\ref{ma24}) and (\ref{GBany5}) are given by 
\be
\label{ma28}
\phi=f^{-1}(t)\quad \left(t=f(\phi)\right)\, ,\quad a=a_0\e^{g(t)}\ \left(H= g'(t)\right)\, .
\ee

We now consider the reconstruction in terms of $N$. By the similar procedure to obtain (\ref{ma26}), we find  
\bea
\label{ma29}
\xi(\phi(N)) &=& \frac{1}{8}\int^N dN_1 \frac{\e^{N_1}}{H(N_1)^3} \int^{N_1} \frac{dN_2}{\e^{N_2}}
\left(\frac{2}{\kappa^2}H' (N_2) + H(N_2) {\phi'(N_2)}^2 \right)\, , \nn
V(\phi(N)) &=& \frac{3}{\kappa^2}H(N)^2 - \frac{1}{2}H(N)^2 \phi' (N)^2 \nn
&& - 3\e^N H(N) \int^N \frac{dN_1}{\e^{N_1}}
\left(\frac{2}{\kappa^2} H' (N_1) + H(N_1) \phi'(N_1)^2 \right)\, .
\eea
Then by using functions $h(\phi)$ and $\tilde f(\phi)$, if $V(\phi)$ and $\xi(\phi)$ are given by
\bea
\label{ma30}
V(\phi) &=& \frac{3}{\kappa^2}h \left(\tilde f(\phi)\right)^2 
 - \frac{h\left(\tilde f\left(\phi\right)\right)^2}{2\tilde f'(\phi)^2} \nn
&& - 3h\left(\tilde f(\phi)\right) \e^{\tilde f(\phi)} \int^\phi d\phi_1 \tilde f'( \phi_1 )
\e^{-\tilde f(\phi_1)} \left(\frac{2}{\kappa^2}h'\left(\tilde f(\phi_1)\right) 
+ \frac{h'\left(\tilde f\left(\phi_1\right)\right)}{f'(\phi_1 )^2} \right)\, ,\nn
\xi(\phi) &=& \frac{1}{8}\int^\phi d\phi_1 
\frac{\tilde f'(\phi_1) \e^{\tilde f(\phi_1)} }{h\left(\tilde f\left(\phi_1\right)\right)^3} \nn
&& \times \int^{\phi_1} d\phi_2  f'(\phi_2) \e^{-\tilde f(\phi_2)} \left(\frac{2}{\kappa^2}h'\left(\tilde f(\phi_2)\right) 
+ \frac{h\left(\tilde f(\phi_2)\right)}{\tilde f'(\phi_2)^2} \right)\, ,
\eea
we find the following solution:
\be
\label{ma31}
\phi=\tilde f^{-1}(N)\quad \left(N=\tilde f(\phi)\right)\, ,\quad H = h(N) \, .
\ee

\section{Einstein-$F(\mathcal{G})$-gravity}

In the action (\ref{ma22}), there appears a scalar field $\phi$.  
The propagation of the scalar field often generates problems in the tests of the Newton law, 
solar system test, etc. 
An easy way to avoid these problems is to drop the kinetic term of the scalar field.
Then we obtain
\be
\label{ma32}
S=\int d^4 x \sqrt{-g}\left[ \frac{R}{2\kappa^2}  
 - V(\phi) - \xi(\phi) \mathcal{G}\right]\, .
\ee
By the variation of the scalar field $\phi$, we obtain an algebraic equation: 
\be
\label{ma33} 
0=V'(\phi) + \xi'(\phi) \mathcal{G} \, ,
\ee
which could be solved, at least locally, with respect to $\phi$ as $\phi= \phi(\mathcal{G})$. 
Then by inserting the obtained expression of $\phi(\mathcal{G})$ into the action (\ref{ma32}), we find
\be
\label{33b}
S=\int d^4 x \sqrt{-g}\left[ \frac{R}{2\kappa^2} + F(\mathcal{G}) \right]\, .
\ee
Here
\be
\label{33c}
F(\mathcal{G}) \equiv - V\left(\phi(\mathcal{G})\right) + \xi\left(\phi(\mathcal{G})\right)\mathcal{G}\, .
\ee
We call the gravity whose action is given by (\ref{33b}) 
as the Einstein-$F(\mathcal{G})$-gravity \cite{Nojiri:2005jg}. 
The comparative review of different modified gravities 
(including scalar-GB, or $F(\mathcal{G})F(G)$) is done in \cite{Nojiri:2006ri}.

In the following, we use the results obtained in \cite{Nojiri:2005jg,Nojiri:2005am,Cognola:2006eg} and consider 
the reconstruction of scalar-$F(\mathcal{G})$-gravity. 
By the variation of the metric tensor in the FRW universe (\ref{ma1}), we obtain the following equations: 
\bea
\label{ma34}
0&=& - \frac{3}{\kappa^2}H^2 + V(\phi) + 24 H^3 \frac{d \xi (\phi(t))}{dt}\, ,\nn
0&=& \frac{1}{\kappa^2}\left(2\dot H + 3 H^2 \right) - V(\phi)
 - 8H^2 \frac{d^2 \xi (\phi(t))}{dt^2} - 16H \dot H \frac{d\xi (\phi(t))}{dt} \nn
&& - 16 H^3 \frac{d \xi (\phi(t))}{dt}\, .
\eea
By using (\ref{ma34}), we obtain
\bea
\label{ma35}
&& \xi (\phi(t)) = \frac{1}{8}\int^t dt_1 \frac{a(t_1)}{H(t_1)^2} W(t_1) \, ,\nn
&& V(\phi(t)) = \frac{3}{\kappa^2}H(t)^2 - 3a(t) H(t) W(t) \, ,\quad 
W(t) \equiv \frac{2}{\kappa^2} \int^{t} \frac{dt_1}{a(t_1)} \dot H (t_1) \, .
\eea
Since there is no kinetic term of $\phi$, we can redefine $\phi$ properly and we may identify $\phi$ with 
the cosmological time: $\phi=t$. By the procedures similar to those in the last section, we find that, 
if we consider the following $V(\phi)$ and $\xi$ given in term of a single function $g$, 
\bea
\label{ma36}
&& V(\phi) = \frac{3}{\kappa^2}g'\left(\phi\right)^2 - 3g'\left(\phi\right) 
\e^{g\left(\phi\right)} U(\phi) \, , \nn
&& \xi (\phi) = \frac{1}{8}\int^\phi d\phi_1 \frac{\e^{g\left(\phi_1\right)} }{g'(\phi_1)^2} U(\phi_1)\, ,\quad 
U(\phi) \equiv \frac{2}{\kappa^2}\int^\phi d\phi_1 \e^{-g\left(\phi_1\right)} g''\left(\phi_1\right) \, ,
\eea
we find the following solution:
\be
\label{ma37} 
a=a_0\e^{g(t)}\ \left(H= g'(t)\right)\, .
\ee

When we use e-foldings $N$ instead of the cosmological time $t$, we obtain
\bea
\label{ma38}
&& \xi (\phi(N)) = \frac{1}{8}\int^N dN_1 \frac{\e^{N_1}}{H(N_1)^3} \tilde W(N_1) \, ,\nn
&& V(\phi(N)) = \frac{3}{\kappa^2}H(N)^2 - 3\e^N H(N) \tilde W(N) \, , \ 
\tilde W(N) \equiv \frac{2}{\kappa^2} \int^{N} \frac{dN_1}{\e^{N_1}} \dot H (N_1) \, .
\eea
Now we identify $\phi$ with the e-folding $N$: $\phi=N$ instead of $\phi=t$. 
Then if $V(\phi)$ and $\xi$ are given by
\bea
\label{ma39}
&& V(\phi) = \frac{3}{\kappa^2}h\left(\phi\right)^2 - 3h\left(\phi\right) 
\e^{\phi} \tilde U(\phi) \, , \nn
&& \xi (\phi) = \frac{1}{8}\int^\phi d\phi_1 \frac{\e^{\phi_1} }{h(\phi_1)^3} \tilde U(\phi_1)\, ,\quad
U(\phi) \equiv \frac{2}{\kappa^2}\int^\phi d\phi_1 \e^{-\phi_1} h'\left(\phi_1\right) \, ,
\eea
we obtain the solution $H=h(N)$. 

\section{$F(R)$-gravity}

$F(R)$-gravity is a modification of Einstein gravity, where the scalar curvature $R$ of the 
Einstein-Hilbert action is replaced by an appropriate function $F(R)$ as 
\be
\label{ma39c}
S=\int d^4 x \sqrt{-g} F(R)\, .
\ee
As a first model of the dark energy in $F(R)$-gravity, the following action was proposed 
 \cite{Capozziello:2002rd,Carroll:2003wy} (see also \cite{Nojiri:2003ft}), 
\be
\label{ma34b}
F(R)=\frac{R}{2\kappa^2} - \frac{\mu^4}{R}\, .
\ee
Now we consider the reconstruction of the $F(R)$-gravity 
based on \cite{Nojiri:2006gh,Nojiri:2006jy,Nojiri:2009kx}. 

First we rewrite the action (\ref{ma39c}) in the following form: 
\be
\label{ma35b}
S=\int d^4 x \sqrt{-g} \left\{P(\phi) R + Q(\phi) \right\}\, .
\ee
Here $\phi$ is an auxiliary scalar field and 
$P$ and $Q$ are functions of scalar field $\phi$. 
Then by the variation of $\phi$, we obtain 
\be
\label{ma40}
0=P'(\phi)R + Q'(\phi)\, ,
\ee
which could be solved with respect to $\phi$ as $\phi=\phi(R)$. By substituting 
the obtained expression of $\phi$ into the action (\ref{ma35b}), we obtain the action in the 
form of (\ref{ma39c}), where
\be
\label{ma41}
F(R) = P(\phi(R)) R + Q(\phi(R))\, . 
\ee

By the variation of the action (\ref{ma39c}) under the metric, we obtain
\be
\label{ma42}
0 = -\frac{1}{2}g_{\mu\nu}\left\{P(\phi) R + Q(\phi) \right\} - R_{\mu\nu} P(\phi)  
+ \nabla_\mu \nabla_\nu P(\phi) - g_{\mu\nu} \nabla^2 P(\phi) \, ,
\ee
which could be rewritten in the following form:
\bea
\label{ma43}
0&=&-6 H^2 P(\phi) - Q(\phi) - 6H\frac{dP(\phi(t))}{dt} \, ,\nn
0&=&\left(4\dot H + 6H^2\right)P(\phi) + Q(\phi) + 2\frac{d^2 P(\phi(t))}{dt} + 4H\frac{d P(\phi(t))}{dt} \, .
\eea
By deleting $Q(\phi)$ in (\ref{ma43}), we obtain
\be
\label{ma44}
0=2\frac{d^2 P(\phi(t))}{dt^2} - 2 H \frac{dP(\phi(t))}{dt} + 4\dot H P(\phi) \, . 
\ee
Since there is an ambiguity to redefine scalar field $\phi$ properly, we may identify $\phi$ with the 
cosmological time $t$: $\phi=t$
Then by assuming the scale factor $a$ is given by a function $g(t)$: $a=a_0\e^{g(t)}$, 
we obtain the following differential equation
\be
\label{ma47}
0=2 \frac{d^2 P(\phi)}{d\phi^2} - 2 g'(\phi) \frac{dP(\phi))}{d\phi} + 4g''(\phi) P(\phi) \, .
\ee
By solving the differential equation, we can find $P(\phi)$ and also $Q(\phi)$:  
\be
\label{ma48}
Q(\phi)=-6 \left(g'(\phi)\right)^2 P(\phi) 
 - 6g'(\phi) \frac{dP(\phi)}{d\phi} \, .
\ee
In principle, almost all arbitrary development of universe can be described by $F(R)$-gravity. 

We now consider an example which reproduces the  $\Lambda$CDM-era without real matter. 
In the Einstein gravity, when there exists a cosmological constant and one kind of matter 
with constant EoS parameter $w$, we find the following behavior of the scale factor $a$: 
\be
\label{ma49}
a = a_0\e^{g(t)}\, ,\quad 
g(t) = \frac{2}{3(1+w)}\ln \left(\alpha \sinh \left(\frac{3(1+w)}{2l}\left(t - t_0 \right)\right)\right)
\, . 
\ee
Here $a_0$ and $t_0$ are constants and $\alpha$ is defined by
\be
\label{ma45}
\alpha^2\equiv \frac{1}{3}\kappa^2 l^2 \rho_0 a_0^{-3(1+w)}\, .
\ee
Here $l$ is the length scale given by the cosmological constant. 
By following the procedures given in this section, we obtain the following differential equation:
\bea
\label{ma50}
0 &=& \frac{d^2 P(\phi)}{d\phi^2} - \frac{1}{l}\coth \left(\frac{3(1+w)}{2l}
\left(t - t_0 \right)\right) \frac{dP(\phi)}{d\phi} \nn
&& - \frac{3(1+w)}{l^2} \sinh^{-2} \left(\frac{3(1+w)}{2l}\left(t - t_0 \right)\right) P(\phi)\, .
\eea
By changing the variable $\phi$ to $z$ defined by 
\be
\label{ma51}
z\equiv - \sinh^{-2} \left(\frac{3(1+w)}{2l}\left(t - t_0 \right)\right) \, ,
\ee
we can rewrite the equation (\ref{ma50}) to 
Gauss' hypergeometric differential equation: 
\bea
\label{ma52}
&& 0=z(1-z)\frac{d^2 P}{dz^2} + \left[\tilde\gamma 
 - \left(\tilde\alpha + \tilde \beta + 1\right)z\right] \frac{dP}{dz}
 - \tilde\alpha \tilde\beta P\, , \nn
&& \tilde\gamma \equiv 4 + \frac{1}{3(1+w)} \, ,\quad
\tilde\alpha + \tilde\beta + 1 \equiv 6 + \frac{1}{3(1+w)}\, ,\quad 
\tilde\alpha \tilde\beta \equiv - \frac{1}{3(1+w)}\, .
\eea
Then the solution is given by the hypergeometric function as follows
\be
\label{ma53}
P= P_0 F(\tilde\alpha,\tilde\beta,\tilde\gamma;z) 
\equiv P_0 \frac{\Gamma(\tilde\gamma)}{\Gamma(\tilde\alpha) \Gamma(\tilde\beta)}
\sum_{n=0}^\infty \frac{\Gamma(\tilde\alpha + n) \Gamma(\beta + n)}{\Gamma(\tilde\gamma + n)}
\frac{z^n}{n!}\, .
\ee
Here $\Gamma$ is the $\Gamma$-function. We also obtain the explicit form of $Q(\phi)$ as 
\be
\label{m34b}
Q = - \frac{6(1-z)P_0}{l^2}F( \tilde\alpha,\tilde\beta,\tilde\gamma;z) 
 - \frac{3(1+w) z(1-z)P_0}{l^2(13 + 12w)}
F(\tilde\alpha+1,\tilde\beta+1,\tilde\gamma+1;z)\, .
\ee

We now consider the reconstruction without auxiliary field and by using $N$ 
based on \cite{Nojiri:2009kx}. 
The equation corresponding to the 1st FRW equation has the following form:
\be
\label{m35b}
0 = - \frac{F(R)}{2} + 3 \left( H^2 + \dot H\right) F'(R) 
 - 18 (\left(4 H^2 \dot H + H\ddot H\right) F''(R) \, .
\ee
Here the scalar curvature $R$ is given by $R = 6\dot H + 12 H^2$. 
By using $N$, Eq.(\ref{m35b}) can be rewritten as
\be
\label{m36}
0 = - \frac{F(R)}{2} + 3 \left( H^2 + H H'\right) F'(R) 
 - 18 (\left(4 H^3 H' + H^2 \left(H'\right)^2 + H^3 H''\right) F''(R) \, .
\ee
We now assume $H$ is given in terms of $N$ or redshift $z$: 
$H=g(N) = g \left(- \ln\left(1+z\right)\right)$. 
Then the scalar curvature is given by $\Rightarrow$ $R = 6 g'(N) g(N) + 12 g(N)^2$, 
which could be solved with respect to $N$: $N=N(R)$. Then we obtain from (\ref{m36})
\bea
\label{m37}
0 &=& -18 \left(4g\left(N\left(R\right)\right)^3 g'\left(N\left(R\right)\right) 
+ g\left(N\left(R\right)\right)^2 g'\left(N\left(R\right)\right)^2 \right. \nn
&& \left. + g\left(N\left(R\right)\right)^3g''\left(N\left(R\right)\right)\right) \frac{d^2 F(R)}{dR^2} \nn
&& + 3 \left( g\left(N\left(R\right)\right)^2
+ g'\left(N\left(R\right)\right) g\left(N\left(R\right)\right)\right) \frac{dF(R)}{dR} - \frac{F(R)}{2} \, ,
\eea
which is the differential equation for $F(R)$, where the variable
is scalar curvature $R$.
If we define $G(N)$ by $G(N) \equiv g\left(N\right)^2 = H^2$ $\Rightarrow$, the equation (\ref{m37}) 
can be little bit simplified as 
\bea
\label{m38}
0 &=& -9 G\left(N\left(R\right)\right)\left(4 G'\left(N\left(R\right)\right)
+ G''\left(N\left(R\right)\right)\right) \frac{d^2 F(R)}{dR^2} \nn
&& + \left( 3 G\left(N\left(R\right)\right)
+ \frac{3}{2} G'\left(N\left(R\right)\right) \right) \frac{dF(R)}{dR} - \frac{F(R)}{2}\, ,
\eea
with $R = 3 G'(N) + 12 G(N)$. 

We now again consider the reconstruction of $F(R)$-gravity reproducing $\Lambda$CDM-era 
without real matter, as an example. 
In the Einstein gravity, the FRW equation has the following form: 
\be
\label{m39b}
\frac{3}{\kappa^2} H^2 = \frac{3}{\kappa^2} H_0^2 + \rho_0 a^{-3}
= \frac{3}{\kappa^2} H_0^2 + \rho_0 a_0^{-3} \e^{-3N} \, .
\ee
Here $H_0$ and $\rho_0$ are constants. Then we obtain the form of $G(N)$ and $R$ as follows:
\be
\label{m39c}
G(N) = H_0^2 + \frac{\kappa^2}{3} \rho_0 a_0^{-3} \e^{-3N} \, , \quad 
R = 12 H_0^2 + \kappa^2\rho_0 a_0^{-3} \e^{-3N}\, .
\ee
The second equation in (\ref{m39c}) can be solved with respect to $N$ as 
\be
\label{m39d}
N = - \frac{1}{3}\ln \left(\frac{ \left(R - 12 H_0^2\right)}{\kappa^2 \rho_0 a_0^{-3}}\right)\, ,
\ee
and we obtain the following differential equation:
\be
\label{m39e}
0 = 3\left(R - 9H_0^2\right)\left(R - 12H_0^2\right) \frac{d^2 F(R)}{d^2 R} 
 - \left( \frac{1}{2} R - 9 H_0^2 \right) \frac{d F(R)}{dR} - \frac{1}{2} F(R)\, .
\ee
By defining the new variable $x=\frac{R}{3H_0^2} - 3$, 
we obtain the hypergeometric differential equation:
\bea
\label{m40}
&& 0=x(1-x)\frac{d^2 F}{dx^2} + \left(\gamma - \left(\alpha + \beta + 1\right)x\right)\frac{dF}{dx}
 - \alpha \beta F\, ,\nn
&& \gamma = - \frac{1}{2}\, ,\alpha + \beta = - \frac{1}{6}\, ,\quad \alpha\beta = - \frac{1}{6}\, ,
\eea
and the solution is given by Gauss' hypergeometric function:
\be
\label{m41}
F(x) = A F(\alpha,\beta,\gamma;x) + B x^{1-\gamma} F(\alpha - \gamma + 1,  \beta - \gamma + 1,
2-\gamma;x)\, .
\ee
Here $A$ and $B$ are constants 

\section{$F(R)$, $F(\mathcal{G})$ v.s. $P(\phi)R + Q(\phi)$, $-V(\phi) + \xi(\phi) \mathcal{G}$ }

This section is based on \cite{Nojiri:2009xh}. 

We start with the following problem: 
The Einstein gravity coupled with a perfect fluid with constant EoS parameter $w$ can be 
reproduced by the following form of $F(R)$:
\be
\label{m44}
F(R) \propto R^m\, , \quad 
w = - 1 - \frac{2(m-2)}{3(m-1)(2m-1)}\, .
\ee
This form of $F(R)$ includes the models with Big Bang $w_\mathrm{BB}>0$ and Big Rip $w=w_\mathrm{BR} <0$ singularities. 
In both case of the singularities, $R$ diverges and goes to infinity. Then if we try to construct a model which shows 
both of the Big-Bang and Big-Rip singularities, $F(R)$ must be a double valued function of $R$.
There is a similarly problem even in $F(\mathcal{G})$-gravity \cite{Goheer:2009qh}. 
$F(G)$ often becomes a double valued function or it may become purely imaginary function when 
we try to construct a model which shows the transition from decelerating expansion to accelerating expansion.     

The actions of $F(R)$-gravity (\ref{ma39c}) and $F(\mathcal{G})$-gravity (\ref{33b}) can be rewritten in the 
form of $P(\phi)R + Q(\phi)$-form (\ref{ma35b}) and $-V(\phi) + \xi(\phi) \mathcal{G}$-form (\ref{ma32}). 
The forms are locally equivalent to each other but as we show now,  
$P(\phi)R + Q(\phi)$-form and $-V(\phi) + \xi(\phi) \mathcal{G}$-form contain wider class of theories than 
$F(R)$-form in (\ref{ma39c}) and $F(\mathcal{G})$-form in (\ref{ma32}). 
To show this, we consider the following toy model in $F(R)$-gravity in $P(\phi)R + Q(\phi)$-form, 
where $P(\phi)$ and $Q(\phi)$ are given by
\be
\label{m45}
P(\phi) = \frac{1}{3} \phi^3 + \beta \phi^2 \, , \quad Q(\phi) =  \gamma \phi \, .
\ee
Here $\beta$ and $\gamma$ are constants. 
By the variation of $\phi$, we obtain
\be
\label{m45b}
0 = \phi^2 + 2 \beta \phi + \gamma R \, ,
\ee
which can be solved with respect to $\phi$ and we obtain
\be
\label{m45c}
\phi = - \beta \pm \sqrt{ \beta^2 - \gamma R}\, ,
\ee
and we find the following expression of the action: 
\be
\label{m45d}
\tilde S_{F(R)} = \int dx^4 \sqrt{-g} \frac{F_\pm (R)}{2\kappa^2}\, ,\quad 
F_\pm (R) \equiv \left( - \frac{2\beta^2}{3} 
+ \frac{\gamma R}{3} \right) \left(  - \beta \pm \sqrt{ \beta^2 - \gamma R} \right)\, .
\ee
Then the action becomes double-valued function and the value of $R$ is restricted to be $\gamma R< \beta^2$ 
for the action to be real.
Then the model in the $F(R)$ form can only describe the theory corresponding to one of 
double-valued function and $R$ is restricted to be $\gamma R< \beta^2$. 
We need not, however, to start from $F(R)$-form but from $P(\phi) R + Q(\phi)$-form. 
Then we may obtain a model with transition between $F_+(R)$ and $F_-(R)$ and/or the region $\gamma R< \beta^2$ 
is not prohibited. 

As a more concrete and more realistic example, we consider the following model \cite{Bamba:2008hq}:
\bea
\label{m46}
&& P(\phi) = e^{\tilde{g}(\phi)/2} \tilde{p}(\phi)\, ,\quad 
Q(\phi) = -6 \left[ \frac{d \tilde{g}(\phi)}{d\phi} \right]^2 P(\phi) 
-6\frac{d \tilde{g}(\phi)}{d\phi} \frac{dP(\phi)}{d\phi} \, ,\nn
&& \tilde{g}(\phi) = - 10 \ln \left[ \left(\frac{\phi}{t_0}\right)^{-\gamma} 
 - C \left(\frac{\phi}{t_0}\right)^{\gamma+1} \right]\, , \quad 
\tilde{p}(\phi) = \tilde{p}_+ \phi^{\beta_+} + \tilde{p}_- \phi^{\beta_-}\, ,\nn
&& \beta_\pm = \frac{1 \pm \sqrt{1 + 100 \gamma (\gamma + 1)}}{2}\, .
\eea
Here $C$ and $\gamma$ are constants. 
Then $P(\phi)$ and $Q(\phi)$ are smooth functions of $\phi$ as long as 
$0<\phi< t_s \equiv t_0 C^{-1/(2\gamma + 1)}$.
An exact solution of this model is given by
\be
\label{m47}
H(t) = \left(\frac{10}{t_0}\right) \left[ \frac{ \gamma \left(\frac{t}{t_0}\right)^{-\gamma-1 }
 + (\gamma+1) C \left(\frac{t}{t_0}\right)^{\gamma} }{\left(\frac{t}{t_0}\right)^{-\gamma}
 - C \left(\frac{t}{t_0}\right)^{\gamma+1}}\right] \, .
\ee
Then when $t\to 0$, i.e., $t \ll t_s$, $H(t)$ behaves as $H(t) \sim \frac{10\gamma}{t}$, 
which corresponds to the Big Bang singularity. 
On the other hand, when $t\to t_s$, $H(t)$ behaves as $H(t) \sim \frac{10}{t_s - t}$, 
which corresponds to the Big Rip singularity. 
Then we can obtain the model describing both of the Big Bang 
and Big Rip singularities in $P(\phi) R + Q(\phi)$ form but not in $F(R)$ form. 

We also consider the following $F(G)$ gravity model:
\bea
\label{m48}
V(\phi) &=& \frac{3}{\phi_0 \kappa^2}\left( 1 + g_1\frac{\phi_0}{\phi} \right)^2 
 - \frac{6g_1}{\phi_0^2 \kappa^2}\left( 1 + g_1\frac{\phi_0}{\phi} \right)
\left(\frac{\phi}{\phi_0}\right)^{g_1} W\left( - g_1 - 1, \frac{\phi}{\phi_0} \right)\, ,\nn
\xi(\phi) &=& \frac{\phi_0^2 g_1}{4\kappa^2} \int^{\frac{\phi}{\phi_0}} dx 
\frac{\e^x x^{g_1}}{\left( 1 + g_1 x \right)^2} W\left( - g_1 - 1, x \right)\, .
\eea
Here $g_1$ and $\phi_0$ are positive constants and $W(\alpha,x)$ is given in terms of the incomplete $\Gamma$ function:
\be
\label{m49}
W(\alpha,x) = \int^x dy \e^{-y} y^{\alpha - 1}\, . 
\ee
We should note that $V(\phi)$ and $\xi(\phi)$ are smooth functions as long as $\phi>0$. 
An exact solution of (\ref{m48}) is given by
\be
\label{m50}
H(t) = \frac{1}{\phi_0} + \frac{g_1}{t}\, .
\ee
When $t$ is small, $H(t)$ behaves as Big Bang singularity and when $t$ is large, 
$H$ goes to a constant, which is asymptotically de Sitter universe. 
Then we can explicitly construct a model which admits a transition from decelerating phase 
to the accelerating phase.

\section{Summary}

We have reviewed the reconstruction in the several gravity models like the Einstein gravity coupled 
with generalized perfect fluid, scalar-Einstein gravity, scalar-Einstein-Gauss-Bonnet gravity, 
Einstein-$F(\mathcal{G})$-gravity, and $F(R)$-gravity. 
Very explicit formulas have been given to reconstruct the models, which could be used when we find 
the detailed data of the development of the universe by future observations. 
Especially we find the formulas using e-foldings $N$, which has a direct relation with observed redshift. 
In this report, just for simplicity, we neglect the contributions from matters. 
We can find, however, that it is easy to include the contribution if we see the original papers. 
As long as we observe the time development of the Hubble rate $H$, there exists a variety of models 
describing the arbitrary development. 
Constraints coming from the several tests of the Newton law, including the solar system test, 
and several requirements from cosmology like structure formation, the scalar/tensor ratio 
of the inflation, etc., however, already excluded some of the models. 
More detailed observation of universe could tell which model could be a real theory. 

\section*{Acknowledgments}

The author is deeply indebted to Professor C.~Q.~Geng for the invitation and 
sincere hospitality. He also acknowledges to Dr. K.~Bamba and Dr. T.~Araki 
for kind hospitality. 
This report is based on the collaborations with S.~D.~Odintsov 
and the work by the author is supported by Global
COE Program of Nagoya University provided by the Japan Society
for the Promotion of Science (G07).

\end{document}